\documentstyle[amssymb,prl,aps,multicols,epsf]{revtex}

\begin{document}
\title{Scalar Field as Dark Matter in the Universe}
\author{Tonatiuh Matos\footnote{E-mail: tmatos@fis.cinvestav.mx},
F. Siddhartha Guzm\'an\footnote{E-mail: siddh@fis.cinvestav.mx} and L.
Arturo
Ure\~na-L\'opez\footnote{E-mail: lurena@fis.cinvestav.mx}}
\address{Departamento de F\'{\i}sica, \\
Centro de Investigaci\'on y de Estudios Avanzados del IPN,\\
AP 14-740, 07000 M\'exico D.F., MEXICO.\\
}
\date{\today}
\maketitle

\begin{abstract}
We investigate the hypothesis that the scalar field is the dark matter and
the dark energy in the Cosmos, wich comprises about 95
the Universe. We show that this hypothesis explains quite well the recent
observations on type Ia supernovae.
\end{abstract}

\draft

\begin{multicols}{2}  
\narrowtext

There are really few questions in science more interesting than that of 
finding out which is the nature of the matter composing the Universe. It
is  amazing that after so much effort dedicated to such question, 
what is the Universe composed of?, it has not been possible to give a
conclusive answer. From the latest observations, we do know that about
95\% of matter in the Universe is of non baryonic nature. The old belief
that matter in Cosmos is made of quarks, leptons and gauge bosons is being
abandoned due to the recent observations and the inconsistences which
spring out of this assumption \cite{schram}. Now we are convinced on the
existence of an exotic non baryonic sort of matter which dominates the
structure of the Universe, but its nature is until now a puzzle.

Recent observations of the luminosity-redshift relation of Ia Supernovae
suggest that distant galaxies are moving slower than predicted by Hubble's
law, that is, an accelerated expansion of the Universe seems to hold \cite
{perlmutter,riess}. Furthermore, measurements of the Cosmic Background
Radiation and the mass power spectrum also suggest that the Universe has the
preferable value $\Omega _{0}=1.$ There should exist a kind of missing
anti-gravitational matter possessing a negative pressure $p/\rho
=\omega <0$ \cite{ostriker} which should overcome the enormous gravitational
forces between galaxies. Moreover, the interaction with the rest of \ the
matter should be very weak to pass unnoticed at the solar system level.
These observations are without doubt among the most important discoveries
of the end of the last century, they gave rise to the idea that the
components of the Universe are matter and vacuum energy $\Omega
_{0}=\Omega _{M}+\Omega _{\Lambda }.$ Models such as the quintessence (a
slow varying scalar field) imply $-1<\omega <0$ and the one using a
cosmological constant, requiring $\omega =-1,$ appear to be strong
candidates to be such missing energy, because both of them satisfy an
equation of state concerning an accelerated behavior of the Universe 
\cite{stein}.

Observations in galaxy clusters and dynamical measurements of the mass in
galaxies indicate that $\Omega _{M}\sim 0.4$, (see for example \cite{turner}%
). Observations of Ia supernovae indicate that $\Omega _{\Lambda }\sim 0.6$ 
\cite{perlmutter,riess}. These observations are in very good concordance
with the preferred value $\Omega _{0}\sim 1.$\ Everything seems to agree.
Nevertheless, the matter component $\Omega _{M}$ decomposes itself in
baryons, neutrinos, etc. and\ dark matter. It is observed that stars
and dust (baryons) represent something like $0.3\%$ of \ the whole matter of
the Universe. The new measurements of the neutrino mass indicate that
neutrinos contribute with about the same quantity as matter. In other 
words, say $%
\Omega _{M}=\Omega _{b}+\Omega _{\nu }+\cdot \cdot \cdot \sim 0.05+\Omega
_{DM}$, where $\Omega _{DM}$ represents the dark matter part of the matter
contributions which has a value $\Omega _{DM}\sim 0.35$. This \ value of
the amount of baryonic matter is in concordance with the limits imposed by
nucleosynthesis (see for example \cite{schram}). But we do not know the
nature neither of the dark matter $\Omega _{DM}$ nor of the dark energy $%
\Omega _{\Lambda };$\ we do not know what is the composition of $\Omega
_{DM}+\Omega _{\Lambda }\sim 0.95$, $i.e.$, the $95\%$ of the whole matter
in the Universe.

In a previous work two of us have shown that the scalar field is a strong
candidate to be the dark matter in spiral galaxies \cite{siddh}. Using the hipothesis that the scalar field is the dark matter in galaxies, we were able to reproduce the rotation curves profile of stars going around spiral galaxies. In fact the scalar potential arising for the explanation of rotation curves of galaxies is exponential. Moreover, by using a Monte Carlo simulation, Hurterer and
Turner have been able to reconstruct an exponential potential for
quintessence which brings the Universe into an accelerating epoch \cite
{turner1}. In this last work there is no explanation for the nature of dark
matter, it is taken the value $\Omega _{DM}\sim 0.35$ without further comments. Recently, there are other papers where the late time attractor solutions for the exponential potential are studied\cite{ferr,maco,barr}. If we are consistent with our previous work, this dark matter should be also
of scalar nature representing the $35\%$ of the matter of the Universe. In
this letter we show that the hypothesis that the scalar field is the
dark matter and the dark energy of the Universe is consistent with Ia
supernovae observations and it could imply that the scalar field is the
dominant matter in the Universe, determining its structure at a
cosmological and at a galactic level. In other words, in this letter we
demonstrate that the hypothesis that the scalar field represents more
than $95\%$ of the matter in the Universe is consistent with the recent
observations on Ia supernovae.

We assume Universe is homogenous and isotropic, so we start with the FRW metric 
\begin{equation}
ds^{2}=-dt^{2}+a^{2}(t)\left[ \frac{dr^{2}}{1-kr^{2}}+r^{2}\left( d\theta
^{2}+\sin ^{2}(\theta )d\phi ^{2}\right) \right]
\end{equation}

The equations governing a Universe with a scalar field $\Phi $ and
a scalar potential $V(\Phi )$ are

\begin{equation}
\ddot{\Phi}+3\frac{\dot{a}}{a}\dot{\Phi}+\frac{dV}{d\Phi }=0,  \label{cphi}
\end{equation}

\begin{equation}
\left( \frac{\dot{a}}{a}\right) ^{2}+\frac{k}{a^{2}}=\frac{\kappa _{o}}{3}%
\left( \rho +\rho _{\Phi }\right)  \label{fried}
\end{equation}

\noindent where $\rho _{\Phi }=\frac{1}{2}\dot{\Phi}^{2}+V(\Phi )$ is the
density of the scalar field, $\rho $ is the density of the baryons, plus
neutrinos, plus radiation, etc, and $\kappa _{o}=8\pi G$. In order to write
the field equations (\ref{cphi}) and (\ref{fried}) in a more convenient
form, we follow \cite{chimen}. We define the function $F(a)$ such that $%
V(\Phi(a) )=F(a)/a^{6}.$ Using the variable $d\eta =1/a^{3}dt$, we can
find a first integral of the field equation (\ref{cphi})

\begin{equation}
\frac{1}{2}\dot{\Phi}^{2}+V(\Phi )=\frac{6}{a^{6}}\int da\frac{F}{a}+\frac{C%
}{a^{6}}=\rho _{\Phi }  \label{firint}
\end{equation}

\noindent being $C$ an integration constant. When the scale factor is
considered as the independent variable, it is possible to integrate the
field equations up to quadratures \cite{chimen}

\begin{eqnarray}
t-t_{0} &=&\sqrt{3}\int {\frac{da}{a\sqrt{\kappa _{0}(\rho _{\Phi }+\rho
)-3k/a^{2}}}}  \label{solll} \\
\Phi -\Phi _{0} &=&\sqrt{6}\int {\frac{da}{a}}\left[ {\frac{\rho _{\Phi
}-F/a^{6}}{\kappa _{0}(\rho _{\Phi }+\rho )-3k/a^{2}}}\right] ^{1/2}
\label{quad}
\end{eqnarray}

In order to compare the data obtained from the Ia supernovae observations
with a scalar field dominated Universe, we write the magnitude-redshift
relation \cite{perlmutter}

\begin{equation}
m_{B}^{effective}=\check{M}_{B}+5\log D_{L}(z;\Omega _{i},\Omega _{\Phi })
\label{mb}
\end{equation}
where $D_{L}=H_{0}d_{L}$ is the ``Hubble-constant-free'' luminosity distance
and $\check{M}_{B}:=M_{B}-5\log H_{0}+25$ is the ``Hubble-constant-free''
B-band absolute magnitude at the maximum of a Ia supernovae. The luminosity
distance $D_{L}$ depends on the model we are working with. In what follows we
compare the observational measurements obtained for $m_{B}^{effective}$ with
a theory defining a scalar field dominated Universe. Using equation (\ref
{firint}), the luminosity distance which depends on the geometry and on the
contents of the Universe in the FRW cosmology (see for example \cite{schmidt}%
), reads for our case

\begin{equation}
d_{l}\left( z;\Omega _{i},\Omega _{\Phi },H_{o}\right) =\frac{(1+z)}{H_{o}%
\sqrt{|k|}}sinn\left(
\sqrt{|k|}\int_{\frac{1}{1+z}}^{1}\frac{dx}{\sqrt{%
U_{\Phi }}}\right)  \label{dl}
\end{equation}
where

\begin{eqnarray}
U_{\Phi } &:&=\left( \sum_{i}\Omega _{i}x^{(1-3w_{i})}\right)
-x^{2}(1-\Omega _{o})  \nonumber \\
&&+\frac{1}{\rho _{c}x^{2}}\left( 6\int dx^{\prime }\frac{F(x^{\prime })}{%
x^{\prime }}+C\right)  \label{Up}
\end{eqnarray}
and

\[
sinn(r)=\left\{ 
\begin{array}{cc}
\sin (r) & (k=+1) \\ 
r & (k=0) \\ 
\sinh (r) & (k=-1)
\end{array}
\right. 
\]

\noindent where $i$ labels for b (baryonic), $\nu $ (neutrinos), r
(radiation), etc. with equations of state $p_{i}=w_{i}\rho _{i}$ for each
component. If we rescale $a_{0}=1$ today, then $x=a=1/(1+z)$, being $z$ the
redshift. Let us now compare the expression (\ref{dl}) with the function
used to fit SNe Ia measurements \cite{per97}, with an equation of state $%
p_{x}=w_{x}\rho _{x}$ for the unknown energy. In this case the luminosity
distance is given by the equation (\ref{dl}) with $U_{X}$ in place of $%
U_{\Phi },$ where

\begin{equation}
U_{X}:=\left( \sum_{i}\Omega _{i}x^{(1-3w_{i})}\right) -x^{2}(1-\Omega
_{o})+x^{(1-3w_{x})}\Omega _{x}.  \label{UX}
\end{equation}

\noindent Observe that both expressions (\ref{Up}) and (\ref{UX}) are very
similar, the only differences are the integral term and the one containing
the constant $C.$ Thus, this comparison extremely suggest that $C=0$ and $%
F(x)=V_{o}x^{s}$, with $V_{0\text{ }}$a constant.

Within a good approximation, we can neglect the present contribution of \
density of baryons, neutrinos etc.,$\ \rho _{om}\ll \rho _{o\Phi }$ because
their contribution represents less than $5\%$ of the matter of the
Universe. The next step is to determine which is the scalar field potential.
Fortunately a flat Universe dominated by scalar field with the function $%
F=V_{0}a^{s}$ has a very important property. We can enunciate this property
in the following theorem:


{\bf Theorem 1.}
{\it Let $\rho _{\Phi }=\frac{6}{a^{6}}\int \frac{F}{a}da$ with $F=V_{0}a^{s}$
in a flat Universe dominated by a scalar field. Then the scalar field
potential $V(\Phi )$, is essentially exponential in the regions where the scalar energy density dominates. }

{\bf Proof:} 
{ \it If the Universe is flat, $k=0$. From equation (\ref{quad}) it
follows that

\begin{equation}
\Phi =\sqrt{\frac{6-s}{\kappa _{0}}}\int {\ \frac{da}{a}\sqrt{\frac{1}{%
1+\left( \frac{\rho _{m}}{\rho _{\Phi }}\right) }.}}
\end{equation}

Thus, if the scalar field dominates ($\rho _{m}\ll \rho _{\Phi }$), this implies $a\simeq \exp (\sqrt{\frac{\kappa _{0}}{6-s}}%
\Phi )$. Then, it follows $V(\Phi )=F(a)/a^{6}\simeq V_{0}\exp (-\sqrt{\kappa
_{0}(6-s)}\Phi )$.}$\blacksquare $


This result strongly states that the scalar potential can only be exponential when the scalar field dominates with no other posibilities like ``power-law'' or ``cosine''.

The theorem fulfills very well the present conditions of the Universe with
the hypothesis we are investigating. Thus, we will take an exponential
potential for the model of the Universe, which implies an extraordinary
concordance with the scalar potential used to explain the rotation curves of
galaxies \cite{siddh}.

With the conditions $C=0$ and $F=V_{0}a^{s},$ equations (\ref{solll}) and (%
\ref{firint}) are easily integrated for a flat Universe. One obtains \cite
{ferr,chimen}

\begin{eqnarray*}
a(t) &=&(K(t-t_{0}))^{\lambda } \\
\Phi -\Phi _{0} &=&\sqrt{\frac{6-s}{\kappa _{0}}}\ln a
\end{eqnarray*}

\noindent where $\lambda =2/(6-s)$. The important quantities obtained from
the solution in terms of the parameter $\lambda $ are: the scalar field and
the scalar potential

\begin{eqnarray}
\Phi (a(t)) &=&\sqrt{\frac{2}{\kappa_o \lambda }}\ln (a)  \label{solution1}
\\
V(\Phi ) &=&V_{o}\exp \left( -\sqrt{\frac{2 \kappa _{o}}{\lambda}}\Phi
\right) ,  \label{solution2}
\end{eqnarray}
the energy density of the scalar field

\begin{eqnarray*}
\rho _{\Phi } &=&\rho _{o\Phi }a^{-\frac{2}{\lambda }} \\
\rho _{o\Phi } &=&\frac{6V_{o}}{6-\frac{2}{\lambda }},
\end{eqnarray*}
the state equation of the scalar field

\[
w_{\Phi }=\frac{2}{3\lambda }-1 
\]
where $p_{\Phi }=\frac{1}{2}\dot{\Phi}^{2}-V(\Phi )=w_{\Phi }\rho _{\Phi }$.
The scale factor

\[
a(t)=\left( \frac{t}{t_{o}}\right) ^{\lambda },
\]
where $t_{o}$ is a normalization constant. The Hubble parameter

\[
H=\frac{\dot{a}}{a}=\lambda t^{-1} 
\]

\noindent and the deceleration parameter

\[
q=-\frac{\ddot{a}}{{\dot{a}^{2}}}a=-\frac{\lambda -1}{\lambda }. 
\]

\begin{figure}[h]
\centerline{ \epsfysize=5cm \epsfbox{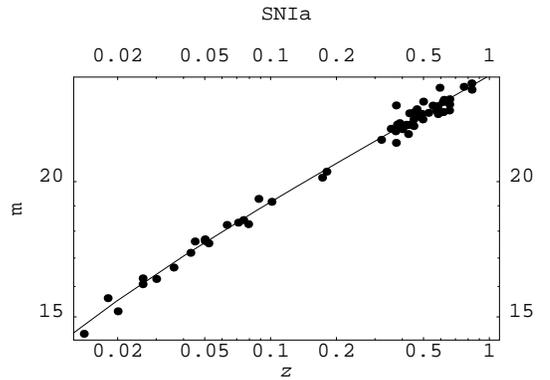}}
\caption{Fit of the solution obtained for the value $\lambda = 1.83$. The
dots represent the observational results and the solid line means $m(z) = 
\check{M} + 5\log D_L$.} 
\label{fig:snia}
\end{figure}

According to the solution (\ref{solution1} - \ref{solution2}), the
expression for the luminosity distance now reads

\begin{equation}
d_{l}\left( z;\lambda ,V_{o},H_{o}\right) =\frac{(1+z)\lambda }{%
H_{o}(1-\lambda )}\left[ \frac{6V_{o}}{s\rho _{c}}\right] ^{-\frac{1}{2}}%
\left[ 1-(1+z)^{\left( 1-\frac{1}{\lambda }\right) }\right]  \label{dist}
\end{equation}

\noindent where $w_{\Phi }=1-s/3$ and we have rescaled $a_{0}=1$ today,
for $\lambda \neq 1$. Fitting (\ref{dist}) with the data of Ia supernovae
\cite{perlmutter,per97} we find $\lambda =1.83$ and $V_{o}=0.78\rho
_{c}$ for $\rho _{0\Phi }\sim 0.95\rho _{c}$ where $\rho _{c}$ is the
critical density ($\rho_c = 0.92 \times 10^{-29} g cm^{-3}$) (see Fig.
\ref{fig:snia}).

Now, we can calculate the deceleration parameter. We obtain $q_{o}=-0.45=constant$, which really implies that the Universe is accelerating. For the density of
the scalar field we obtain $\rho _{\Phi }=0.95\rho _{c}a^{-1.09}$ and for
its equation of state  $w_{\Phi }=-0.636=constant$. Currently, we are 
investigating the CMBR and the mass power spectrum. See\cite{stein} for a 
scalar field with equation of state $w = -2/3$ and $\Omega_{\Phi}$ up to 
$0.8$, where it is concluded that the scalar field fits all the required
observations. If we use $H_o = 70 \frac{Km}{s Mpc}$, we find:

\[
t_{o}=25.6\times 10^{9}yr. 
\]
$t_o$ would be the age of a Universe that was always dominated by the scalar
field, which is not our case.

The great concordance of our hypotheses with experimental results, suggests
that the Universe lies at this moment in a scalar field dominated epoch.
This
permits us to speculate about the behavior of the Universe for red-shifts
greater than $z=1$ as restricted by SNIa observations. Observe that our
results do not imply
that the Universe has been dominated by a scalar field during
all its evolution. Instead, our model accepts the possibility of a Universe
dominated by radiation or matter before the epoch we have analyzed. In order
to draw a complete history of the Universe, we consider the periods of
radiation and matter dominated eras. A general integration of the
conservation equation for a perfect fluid made of radiation (dust),
indicates that the density scales as $\rho _{r}=\rho _{or}a^{-4}$ ($\rho
_{m}=\rho _{om}a^{-3}$), with $\rho _{or}=10^{-5}\rho _{c}$ ($\rho
_{om}=0.05\rho _{c}$). In the FRW standard cosmology, the Universe was
radiation dominated until $a\sim 10^{-3}$, the time when the density of
radiation equals the density of matter. Recalling our result $\rho _{\Phi
}=\rho _{o\Phi }a^{-1.09}$, the Universe changed to be matter dominated
until $a\sim 0.21$, when the density of the scalar field equals the density
of matter. At this time, the density of radiation is negligible. This
corresponds to redshifts $z=3.7$. The implications of this model are very
strong. Since this time (approximately $14\times 10^{9}$ yr. ago for this
model), the scalar field began to dominate the expansion of the Universe and
it enters in its actual acceleration phase, which includes most of the
history of the Universe. Then we wonder if the scalar field is the
responsible for the formation of structure too. According to \cite{madau},
the formation of galaxies started at a few redshifts, from approximately
4.5 to 2, just when the scalar field began to be important.

Some final remarks. With our values, the solution is {singular}, i.e., 
$a(t)$
vanishes at some finite time. Moreover, the solution has no {particle horizon%
} \cite{chimen} as can be seen from the expression (\ref{solll}) because $%
s>4 $. The question why nature uses only spin 1 and spin 2 fundamental 
interactions over the simplest spin 0 interactions becomes clear with our 
result. This result tells us that in fact nature has preferred the spin 0 
interaction over the other two and in such case, the scalar field should
thus be the responsible of the cosmos structure.

\bigskip

\section{Acknowledgements}

We would like to thank Dario Nu\~{n}ez and Michael Reisenberger for many
helpful discussions. This work was partly supported by CONACyT, M\'{e}xico,
under grants 3697-E (T.M.), 94890 (F.S.G.) and 119259 (L.A.U.)

\end{multicols}

\end{document}